\documentclass{jpsj3}
\usepackage{txfonts}

\usepackage{dcolumn}
\usepackage{bm}
\usepackage{color}
\usepackage{float}
\usepackage{siunitx}
\usepackage{physics}

\title{Intrinsic Anomalous Hall Effect Arising from Antiferromagnetism \\
	as Revealed by High-Quality NbMnP}

\author{Yuki Arai$^1$\thanks{222s102s@stu.kobe-u.ac.jp}, Junichi Hayashi$^2$, Keiki Takeda$^2$, Hideki Tou$^1$, Hitoshi Sugawara$^1$, and Hisashi Kotegawa$^1$}
\inst{$^1$Department of Physics, Kobe University, Kobe, Hyogo 657-8501, Japan \\
	$^2$Muroran Institute of Technology, Muroran, Hokkaido 050-8585, Japan
}

\abst{
	The anomalous Hall effect (AHE) in antiferromagnetic (AF) materials arises from symmetry breaking, which is equivalent to that in ferromagnets.
	This suggests that the AHE is induced by the intrinsic mechanism of the band-structure effect, which in turn induces dissipationless transverse conductivity.
	Confirmation of impurity-insensitive anomalous Hall conductivity (AHC) is crucial to this interpretation; however, experimental investigations of the dissipationless AHC for AF materials have been limited by a lack of high-quality samples. In this study, we demonstrated that the AF material NbMnP, which exhibits a large AHE, offers high-quality single crystals. Our findings clearly reveal that the large AHC and small net magnetization of $\sim10^{-3} \mu_{\mathrm{B}}$/Mn are inherent in this material, irrespective of disorder. NbMnP is a novel AF material that generates FM responses in regimes with reduced impurity scattering.
}

\begin{document}
	\maketitle

	Interpretation of the anomalous Hall effect (AHE) has long been controversial in condensed matter physics \cite{nagaosa_anomalous_2010}.
	Karplus and Luttinger proposed that an external electric field induces an anomalous transverse velocity through spin--orbit interactions \cite{KL}.
	Consequently, because it is a band-structure effect, the anomalous Hall conductivity (AHC) is expected to be independent of any impurity scattering; this mechanism is called an intrinsic AHE.
	However, impurity scattering, which is recognized as skew or side-jump scattering, has also been shown to trigger the AHE, that is, an extrinsic AHE \cite{Smit1958,Berger1970}. 
	The absence of an intuitive explanation for the dissipationless intrinsic AHE has prevented it from becoming the mainstream interpretation of AHE; however, in the 2000s, the intrinsic AHE was reconstructed using the Berry phase concept, and first-principles calculations enabled researchers to obtain a theoretical value of the AHC \cite{nagaosa_anomalous_2010,Fang2003,Yao2004,Haldane2004}.
	Currently, there are two widely recognized methods for assessing the origin of AHE.
	The first involves a comparison of the experimental and calculated AHCs \cite{Fang2003,Yao2004}; however, accurate calculations of the AHC under the magnetically ordered states are required.
	The second is to investigate the impurity-scattering dependence of the AHC. Although not related to the validity of the theoretical calculations, relatively clean samples are required because the dissipationless AHC experimentally appears only in the moderate-conductivity regime \cite{Onoda2008}.

	One success introduced by the Berry phase concept was the discovery of a large AHE arising from an antiferromagnetic (AF) structure \cite{Chen14,Nakatsuji2015}, which can occur when the magnetic point group of the AF state allows ferromagnetic (FM) states.
	After the discovery of a large AHE in Mn$_3$Sn \cite{Nakatsuji2015}, several AF materials were shown to exhibit AHE at zero field, accompanied by small net magnetization \cite{Kiyohara16,Nayak16,Ghimire18,Park22}.
	The consistency between the experimental and theoretical values of the AHC presented  in some materials suggests that AF structures induce intrinsic AHE \cite{Nakatsuji2015,Kiyohara16,Nayak16,Kuber,Suzuki17}. 
	However, experimental confirmation of the impurity-scattering dependence is yet to be sufficiently investigated for AF materials because of the difficulty in controlling impurity scattering, owing to a lack of high-quality crystals.
	Specifically, the residual resistivity ratios (RRRs) of AF materials have only reached approximately $2$--$3$ (for Mn$_3$Sn, Mn$_3$Ge, NbCo$_3$S$_6$, and TaCo$_3$S$_6$).
	For $\alpha$-Mn, a crystal with $\mathrm{RRR} = 17$ can be obtained, and the impurity-scattering dependence of the AHC has been confirmed \cite{Akiba20}; however, the magnetic structure under pressure that yields AHE remains unclear. 
	
	\begin{figure}[htb]
		\begin{center}
			\includegraphics[width=0.95\linewidth]{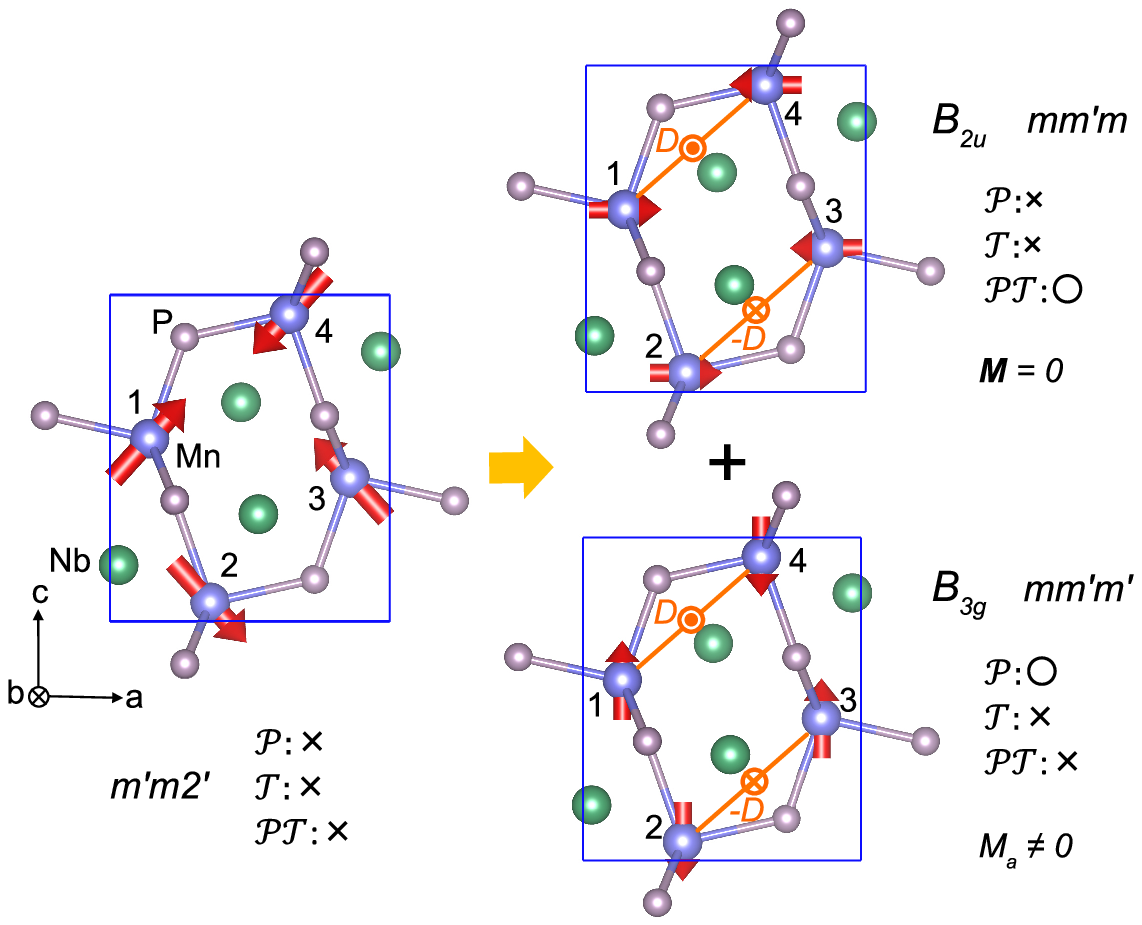}
			\caption{(Color online) The $Q=0$ AF structure of NbMnP, represented by a linear combination of two irreducible representations \cite{Matsuda}. The unit cell includes four Mn atoms labeled Mn1--Mn4, which are equivalent in the paramagnetic state. This combination lowers the symmetry from $Pnma$ to $Pmn2_1$. The actual magnetic point group is $m'm2'$. In the $B_{2u}$ component, $\mathcal{PT}$ symmetry protects the compensated AF structure, even though the Dzyaloshinskii--Moriya (DM) interaction applies. In the $\mathcal{PT}$-symmetry broken $B_{3g}$ component, the DM interaction induces magnetization along the $a$ axis.}
			\label{f1}
		\end{center}
	\end{figure}

	Recently, we identified a large AHE in the noncollinear AF material NbMnP \cite{Kotegawa_NbMnP}.
	The AF structure of NbMnP, illustrated using {\it VESTA} \cite{VESTA}, is represented by a combination of magnetic point groups $mm'm$ (irreducible representation $B_{2u}$) and $mm'm'$ ($B_{3g}$), as shown in Fig.~1 \cite{Matsuda}.
	$mm'm$ ($B_{2u}$) does not possess space-inversion symmetry ($\mathcal{P}$) and time reversal symmetry ($\mathcal{T}$), but instead maintains $\mathcal{PT}$ symmetry.
	In $mm'm'$ ($B_{3g}$), which is represented by a magnetic toroidal quadrupole \cite{Yatsushiro}, $\mathcal{PT}$ symmetry is broken and can induce an AHE from the AF structure because it symmetrically allows FM components along the $a$-axis.
	The observed AHC of $\sigma_{\mathrm{H}} = 230$ $\Omega^{-1}$cm$^{-1}$ is approximately consistent with the theoretical AHC of $\sigma_{\mathrm{H}} = 273$ $\Omega^{-1}$cm$^{-1}$ \cite{Kotegawa_NbMnP}. 
	This agreement supports the intrinsic nature of AHE.
	In contrast, the impurity-scattering dependence of $\sigma_{\mathrm{H}}$ has not been verified, even in NbMnP, because its RRR value was 2, as with other AF materials exhibiting the AHE \cite{Nakatsuji2015,Kiyohara16,Nayak16,Ghimire18,Park22}.
	High-quality single crystals of NbMnP must be used to confirm the origin of the AHE.

	In this Letter, we report single-crystal growth of NbMnP using the Ga-flux method, which differs from the previous self-flux method \cite{Kotegawa_NbMnP}.
	The RRR of the new crystals exceeded 40, which is much higher than the value of 2 obtained in our previous study, indicating a large improvement in the sample quality.
	The high-quality NbMnP exhibited a large AHE, whose AHC was comparable to that of the previously used crystal with a low RRR.
	These results suggest that the AHE in NbMnP arises intrinsically from the band structure of the AF spin configuration.

	\begin{figure}[b]
		\centering
		\includegraphics[width=0.95\linewidth]{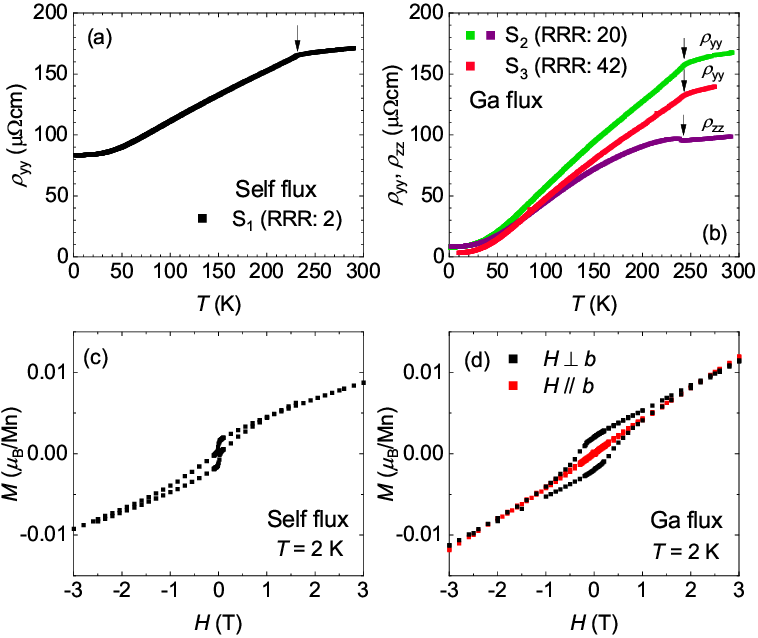}
		\caption{(Color online) Temperature dependence of electrical resistivity for (a) self-flux and (b) Ga-flux crystals. The marked improvement in RRR confirms that the Ga-flux method is effective in obtaining high-quality crystals. Anisotropy between $\rho_{yy}$ and $\rho_{zz}$ was checked for S$_2$, and was weak at low temperatures. Magnetization curves at 2 K for (c) many unoriented self-flux crystals \cite{Kotegawa_NbMnP} and (d) aligned Ga-flux crystals. The similar spontaneous magnetization irrespective of the sample quality suggests that they are inherent to NbMnP.}
		\label{f2}
	\end{figure}

	For the Ga-flux method, a mixture of Nb, Mn, P, and Ga at a respective molar ratio of $1:1:1:30$ was placed in an Al$_2$O$_3$ crucible and sealed in an evacuated quartz ampoule.
	The ampoule was gradually heated to $1050\,{\rm ^o C}$ and maintained at this temperature for 72 h, followed by a slow cooling to $650\,{\rm ^o C}$ at $-5\,{\rm ^o C/h}$. 
	Needle-like single crystals were obtained via centrifugation.
	The crystal symmetry and lattice parameters were determined by single-crystal X-ray diffraction measurements using a Rigaku Saturn724
	diffractometer with multilayer mirror monochromated Mo-K$_{\alpha}$ radiation at room temperature.
	We measured both electrical and Hall resistivities using a standard four-probe method and antisymmetrized the Hall resistivity against magnetic fields to remove the longitudinal component induced by contact misalignment.
	Furthermore, we measured the magnetic properties using a commercial SQUID magnetometer and fixed several crystal pieces.

	\begin{table}[htb]
		\caption{Structural parameters of NbMnP for two crystals obtained separately by the self-flux \cite{Matsuda} and Ga flux methods,  determined by single-crystal X-ray diffraction measurements at $T=293\, \mathrm{K}$. The Wyckoff positions of all atoms were $4c$, and all clear differences are underlined. The occupancy (Occ.) of the Nb site was improved using the Ga-flux method. }
		\label{t1}
		\vspace{1ex}
		\raggedright
		\begin{center}
			\begin{tabular}{cccc}
				\hline
				NbMnP & self-flux & 293 K & \\
				\hline
				Nb & Mn & P \\
				\hline
				$x$ & 0.03102(5) & 0.14147(9) & 0.26798(15) \\
				$y$ & 0.25000 & 0.25000 & 0.25000 \\
				$z$ & 0.67215(5) & 0.05925(8) & 0.36994(13) \\
				Occ. & \underline{0.968} & 1 & 1 \\
				$U$(\AA$^2$) & 0.00582(15) & 0.0063(2) & 0.0061(2) \\
				\hline
			\end{tabular}\\
			\vspace{1ex}
			Orthorhombic ($Pnma$) \\
			$a$=6.1823(2) \AA, \underline{$b$=3.5573(2) \AA}, \underline{$c$=7.2187(3) \AA}, $R$=1.90\% \\
			\vspace{5ex}
			\begin{tabular}{cccc}
				\hline
				NbMnP & Ga flux & 293 K & \\
				\hline
				Nb & Mn & P \\
				\hline
				$x$ &0.03122(7) & 0.14162(12) & 0.2682(2) \\
				$y$ & 0.25000 & 0.25000 & 0.25000 \\
				$z$ &0.67214(6) & 0.05941(10) & 0.37019(19) \\
				Occ. & \underline{0.995} & 1 & 1 \\
				$U$(\AA$^2$) & 0.0052(3) & 0.0059(4) & 0.0059(4) \\
				\hline
			\end{tabular}\\
			\vspace{1ex}
			Orthorhombic ($Pnma$) \\
			$a$=6.1899(2) \AA, \underline{$b$=3.5478(1) \AA}, \underline{$c$=7.2380(2) \AA}, $R$=3.18\% \\
			\vspace{5ex}
		\end{center}
	\end{table}

	Figures \ref{f2}(a) and (b) compare the electrical resistivities of different crystals,
	where $x$, $y$, and $z$ correspond to the $a$, $b$, and $c$-axes, respectively.
	$\rho_{yy}$ exhibits kinks at $T_{\rm N}= 233$ K for the self-flux crystal and at 244 K for the Ga-flux crystals. 
	A significant change appears in the residual resistivity: $\mathrm{RRR}=2$ for the self-flux crystal (Sample 1: S$_1$) and $\mathrm{RRR}=42$ for the Ga-flux crystals (Sample 3: S$_3$), clearly indicating that the Ga-flux method is highly effective for obtaining high-quality single crystals of NbMnP. 
	$T_{\mathrm{N}}$ increased by approximately $10$ K, accompanied by an improvement in the sample quality.
	For Sample 2: S$_2$, we also measured $\rho_{zz}$, which exhibited a small jump at $T_{\rm N}$. The anisotropy between $\rho_{yy}$ and $\rho_{zz}$ was significant at high temperatures, whereas it weakened at low temperatures, indicating that the residual resistivity was isotropic.
	Figures \ref{f2}(c) and 2(d) show a comparison of the magnetization curves at 2 K.
	In the self-flux NbMnP with a lower RRR, a weak FM component of approximately $10^{-3} \mu_{\mathrm{B}}$/Mn appeared \cite{Kotegawa_NbMnP} perpendicular to the $b$-axis \cite{Zhao}. 
	The high-quality NbMnP also exhibited an FM component on the same order of $10^{-3} \mu_{\mathrm{B}}$/Mn, indicating that the net magnetization does not originate from poor sample quality, but is intrinsic. 
	This also supports the interpretation that weak net magnetization originates from spin canting through DM interactions \cite{Kotegawa_NbMnP}.
	As shown in Fig.~1, a DM interaction occurs between two Mn atoms through AF coupling.
	The DM vectors for these couplings are directed along the $b$-axis, inducing spin canting in the $ac$ plane because these bonds lie in the mirror plane.
	There is no DM interaction between Mn1 and Mn3 (or Mn2 and Mn4) because of the inversion symmetry of the crystal.
	Spin canting through DM interactions does not yield net magnetization in the $B_{2u}$ representation with $\mathcal{PT}$ symmetry, because the AF couplings between Mn1 and Mn3 (or Mn2 and Mn4) are protected.
	For the $\mathcal{PT}$-symmetry broken $B_{3g}$ representation, the DM interaction induces nonzero net magnetization along the $a$-axis.

	We then evaluated the crystallographic differences between the two crystal types using single-crystal X-ray diffraction measurements. 
	The results are summarized in Table I.
	Although the space group remains the same, the length of the $b$-axis slightly decreases and that of the $c$-axis increases, for the Ga-flux crystal.
	A significant difference appears in the occupancy at the Nb site, which indicates a nonnegligible deficiency in the self-flux crystal.
	The deficiency at the Nb site almost disappeared in the Ga-flux crystals, suggesting that the stoichiometry was improved using the Ga-flux method.

	\begin{figure*}[htb]
		\centering
		\includegraphics[width=\linewidth]{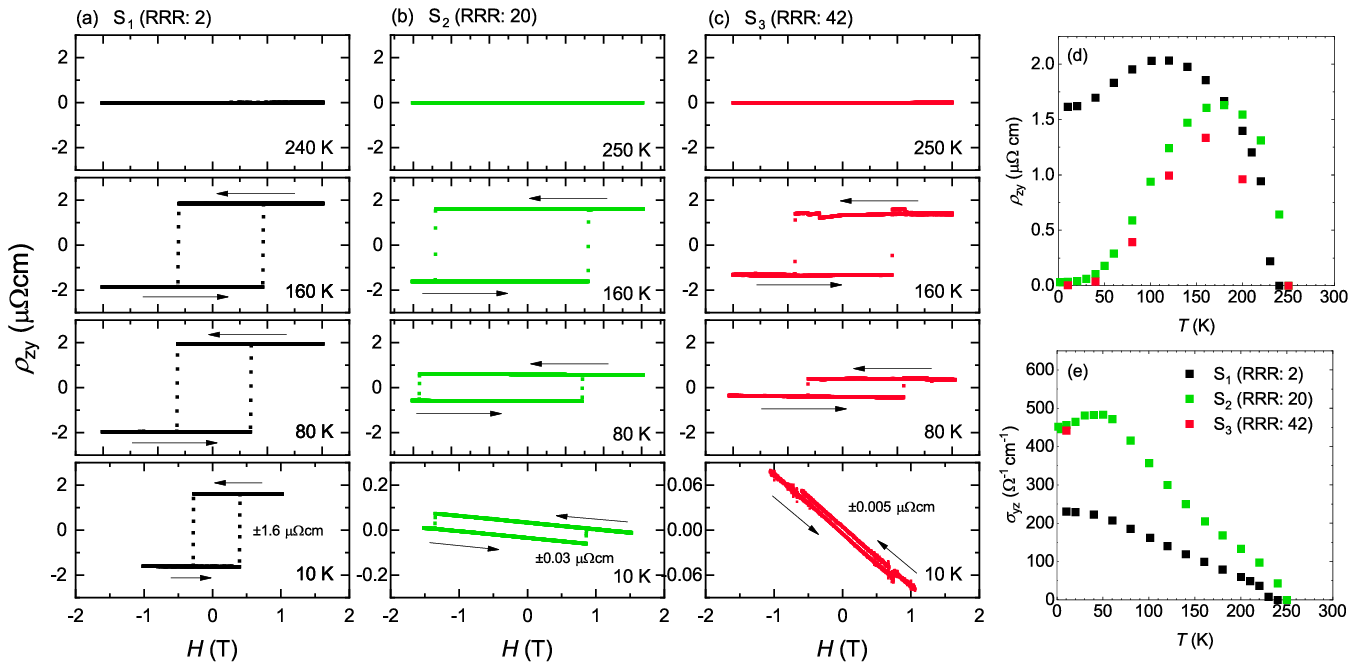}
		\caption{(Color online) (a--c) Field dependence of Hall resistivity $\rho_{zy}$ for crystals with different RRRs. The AHE exhibits hysteresis against the positive and negative magnetic fields and appears below $T_{\mathrm{N}}$ for all crystals. For $\mathrm{RRR}=2$ \cite{Kotegawa_NbMnP}, the large $\rho_{zy}$ remains at low temperatures, while it is suppressed, and the ordinary Hall effect becomes remarkable as RRR increases. For all crystals, measurements were performed after zero-field cooling to suppress the exchange-bias effect of the material \cite{Kotegawa_NbMnP}. (d) Temperature dependence of $\rho_{zy}$ at zero field, i.e., the AHE component. (e) Temperature dependence of the AHC $\sigma_{zy}$, which was converted through $\sigma_{yz} \simeq \rho_{zy}/\rho_{yy}\rho_{zz}$ for S$_2$. For S$_3$ without the $\rho_{zz}$ data, $\sigma_{zy}$ was obtained at low temperatures through $\rho_{yy} \simeq \rho_{zz}$, because the residual resistivity is isotropic. The $\rho_{yy} \simeq \rho_{zz}$ was also assumed for the low-quality S$_1$, where the residual resistivity is large. A two times larger $\sigma_{yz}$ was obtained for the Ga flux (S$_2$ and S$_3$) crystals compared to the self-flux crystal (S$_1$), while these values were similar between $\mathrm{RRR}=20$ and 42. }
		\label{f3}
	\end{figure*}

	Figures 3(a--c) show the Hall resistivity $\rho_{zy}$ for three crystals with $\mathrm{RRR}=2$, 20, and 42.
	Magnetic fields were applied along the $a$-axis to align the AF domains. 
	The emergence of hysteresis below $T_{\rm N}$ indicates that all the crystals exhibited zero-field AHE in the magnetically ordered state. 
	The positive (negative) magnetic field selects the AF domain that induces a positive (negative) $\rho_{zy}$ and generates hysteresis.
	At 160 K, the magnitudes of all $\rho_{zy}$ were similar among the three crystals, whereas they were strongly suppressed toward low temperatures in the high-quality crystal, where the field-dependent ordinary Hall effect was remarkable compared to the AHE.
	Although both the width of the hysteresis and the exchange-bias effect \cite{Kotegawa_NbMnP} depend on the crystal, the relationship remains unclear.
	The suppression of $\rho_{zy}$ was clearly observed in the temperature dependence of $\rho_{zy}$, as shown in Fig.~3(d).
	This behavior is explained by the difference in sample quality, that is, the difference in electrical resistivity.
	In general, $\rho_{zy}$ is related to the Hall conductivity through $\sigma_{yz} \simeq \rho_{zy}/\rho_{yy}\rho_{zz}$.
	When the band structure determines the dissipationless Hall conductivity $\sigma_{yz}$, $\rho_{zy}$ strongly depends on electrical resistivity, thereby reflecting the sample quality.
	For S$_2$, we obtained $\sigma_{yz}$ through $\sigma_{yz} \simeq \rho_{zy}/\rho_{yy}\rho_{zz}$ (Fig.~3(e)).
	The experimental data for the previous S$_1$ are also plotted, where $\rho_{yy} \simeq \rho_{zz}$ was assumed \cite{Kotegawa_NbMnP}.
	This assumption is reasonable because electrical resistivity is dominated by residual resistivity, which is expected to be isotropic. 
	$\sigma_{yz}$ for S$_2$ was found to increase at low temperatures, and its temperature variation was suppressed below approximately $50$ K.
	For S$_3$ without the $\rho_{zz}$ data, we plotted only the low-temperature data by assuming $\rho_{yy} \simeq \rho_{zz}$ because the anisotropy confirmed in S$_2$ was weak at low temperatures.
	Similar $\sigma_{yz}$ were obtained for S$_2$ and S$_3$.
	The values at the lowest temperature were 230 $\Omega^{-1}$cm$^{-1}$ for $\mathrm{RRR}=2$ and $\sim450$ $\Omega^{-1}$cm$^{-1}$ for $\mathrm{RRR}=20$ and 42, thereby demonstrating an approximately twofold difference among the crystals. However, this difference is small compared to the 20 times difference in RRR.

	\begin{figure*}[ht]
		\centering
		\includegraphics[width=\linewidth]{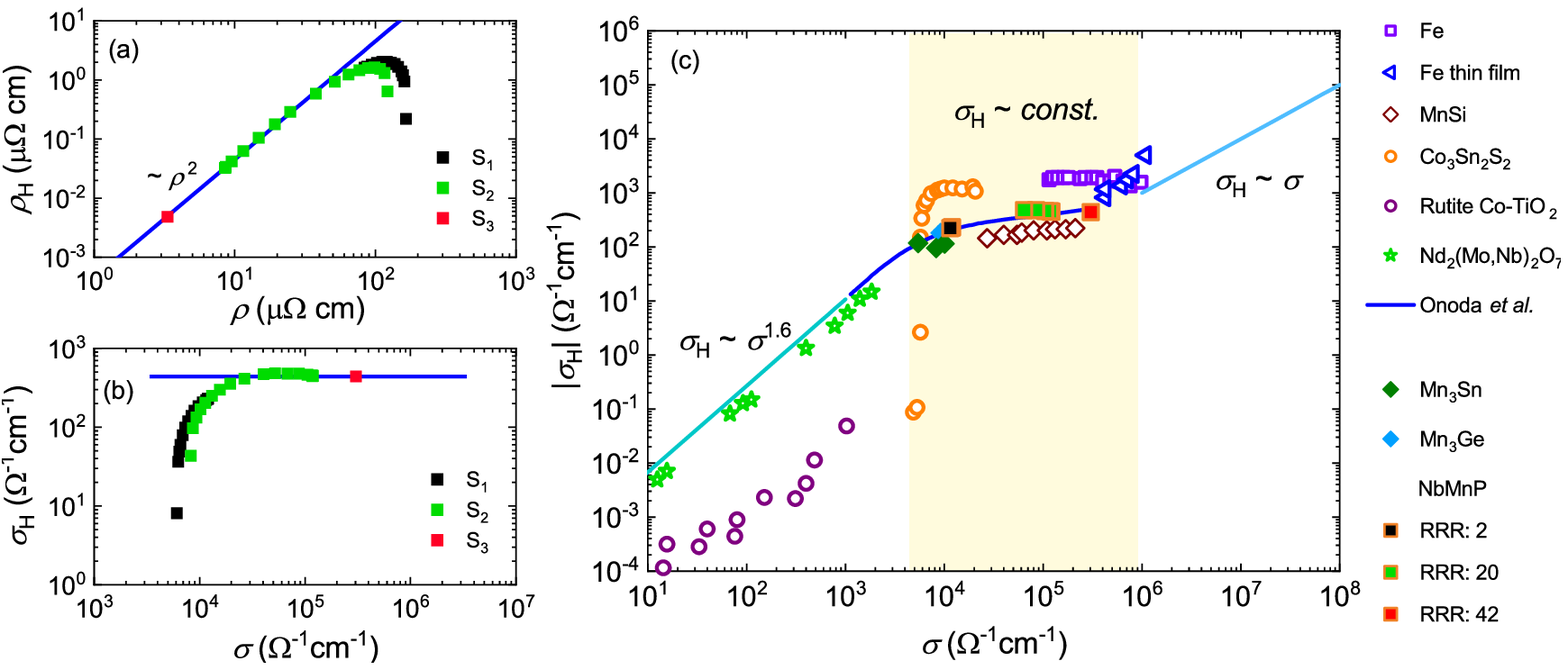}
		\caption{(Color online) Scaling relations in NbMnP between (a) anomalous Hall resistivity $\rho_H$ ($=\rho_{zy}$) and longitudinal electrical resistivity $\rho$, and (b) anomalous Hall conductivity $\sigma_{\mathrm{H}}$ ($=\sigma_{yz}$) and longitudinal conductivity $\sigma$. At temperatures below $\sim50$ K, $\rho_{\mathrm{H}}$ obeys $\sim \rho^2$, and $\sigma_{\mathrm{H}}$ approaches constant values against $\sigma$. These behaviors suggest that the observed AHE mainly arises from the band-structure effect without dissipation. (c) Scaling relation for many ferromagnets \cite{Onoda2008,Miyasato,Lee,Manyala,Toyosaki,Higgins,Liu}, Mn$_3$$X$ ($X=$Sn and Ge) \cite{Chen2021}, and NbMnP. The ferromagnets are represented by open symbols, whereas the AF materials are represented by closed symbols. The data at low temperatures are plotted for NbMnP, showing a weak $\sigma$ dependence in the wide conductivity region. The blue curve indicates the theoretical expectation with an appropriate amount of impurity \cite{Onoda2008}.}
		\label{f3}
	\end{figure*}

	To assess the impurity-scattering dependence of the AHC, we plotted $\rho_{\mathrm{H}}$ ($=\rho_{zy}$) against $\rho$ in Fig.~4(a), and $\sigma_{\mathrm{H}}$ ($=\sigma_{yz}$) against $\sigma$ ($= 1/\rho)$ in Fig.~4(b).
	Here, $\rho = \sqrt{\rho_{yy} \rho_{zz}}$ was used for S$_2$, whereas $\rho = \rho_{yy}$ was applied to S$_1$ and low-temperature data for S$_3$.
	Below approximately 50 K, $\rho_{H}$ follows $\rho^2$; therefore, $\sigma_{H}$ is approximately constant with respect to $\sigma$.
	These relationships demonstrate experimentally that the observed AHE arises through an intrinsic mechanism \cite{nagaosa_anomalous_2010,KL}. 
	This could not be confirmed using the previous low-quality crystal S$_1$ because of the weak temperature dependences of $\rho$ and $\sigma$.
	Figure~4(c) presents $|\sigma_{\mathrm{H}}|$ versus $\sigma$ for several ferromagnets \cite{Onoda2008,Miyasato,Lee,Manyala,Toyosaki,Higgins,Liu}, Mn$_3$$X$ ($X=$Sn and Ge) \cite{Chen2021}, and NbMnP.
	Experimental data for ferromagnets demonstrate that $\sigma_{\mathrm{H}}$ is independent of $\sigma$ in the intermediate-conductivity region of $10^4< \sigma <10^6$ $\Omega^{-1}$cm$^{-1}$ \cite{Onoda2008}.
	In the dirty region of $\sigma < 10^3$--$10^4$ $\Omega^{-1}$cm$^{-1}$, $\sigma_{\mathrm{H}} \sim \sigma^{1.6}$ has been established, whereas $\sigma_{\mathrm{H}} \sim \sigma$ is expected to be dominant in the high-conductivity region owing to skew scattering \cite{Onoda2008}.
	For AF materials, Mn$_3$$X$ remains at approximately $10^4$ $\Omega^{-1}$cm$^{-1}$ in the intermediate region \cite{Chen2021}, whereas NbMnP covers a wide $\sigma$ range of more than one order of magnitude. 
	The weak $\sigma$-dependence of $\sigma_{\mathrm{H}}$ observed on the logarithmic scale resembles the behavior of ferromagnets, ensuring that the AHE under the AF structure is dominated by an intrinsic mechanism in the same framework as that of ferromagnets.

	The prospective benefits of obtaining high-quality NbMnP are not limited to understanding the AHE.
	The AF structure in the FM point group also generates FM responses, such as the anomalous Nernst effect (ANE) \cite{Li17,Ikhlas17}.
	The influence of disorder on the ANE has not been thoroughly investigated, even for ferromagnets; thus, impurity dependence is important to its understanding \cite{Ding}.
	Our preliminary experiment confirmed the emergence of ANE in NbMnP; therefore, it is a suitable material for assessing how impurities affect ANE under the AF spin configuration. 
	Another approach is the elucidation of the asymmetric hysteresis that appears after field cooling in NbMnP \cite{Kotegawa_NbMnP}.
	This exchange-bias effect is usually observed in artificial bilayers of FM and AF materials \cite{Meiklejohn,Nogues}, and a similar behavior has been observed in Co$_3$Sn$_2$S$_2$ \cite{Lachman} and is sample-size dependent \cite{Noah}.
	For NbMnP, we speculate that Nb deficiency affects its asymmetry through the pinning of magnetic moments near the deficiency \cite{Kotegawa_NbMnP}.
	This study showed that the asymmetry still appears, particularly for crystals with ${\rm RRR}=20$.
	Further investigations are required; however, careful disorder dependence is crucial.

	In summary, high-quality NbMnP crystals with ${\rm RRR}>40$ were effectively obtained using the Ga-flux method.
	These crystals showed a small net magnetization and large AHC, which were comparable to those found in previous low-quality crystals, revealing that both effects arise from intrinsic mechanisms.
	NbMnP single crystals, which cover a wide range of electrical conductivities, offer opportunities to systematically investigate the influence of disorder on the FM responses arising from the AF spin configuration.

	\section*{Acknowledgements}
	We thank Michi-To Suzuki for valuable discussions. 
	This work was supported by JSPS KAKENHI Grant Nos. 21K03446 and 23H04871, the Iketani Science and Technology Foundation, and the Murata Science Foundation.

\end{document}